# Independent wavefront tailoring in full polarization channels by helicity-decoupled metasurface

*Guowen Ding, Ke Chen\*, Na Zhang, Junming Zhao, Tian Jiang, and Yijun Feng†*

\*Ke Chen: ke.chen@nju.edu.cn;
†Yijun Feng: yjfeng@nju.edu.cn.

School of Electronic Science and Engineering, Nanjing University, Nanjing, 210093, China

**Abstract:** Controlling the polarization and wavefront of light is essential for compact photonic systems in modern science and technology. This may be achieved by metasurfaces, a new platform that has radically changed the way people engineer wave-matter interactions. However, it still remains very challenging to generate versatile beams with arbitrary and independent wavefronts in each polarization channel by a single ultrathin metasurface. By modulating both the geometric and propagation phases of the metasurface, here we propose a method that can generate an assembly of circularly- and linearly-polarized beams with simultaneously the capability of independent encoding desired wavefront to each individual polarization channel, which we believe will greatly enhance the information capacities of the meta-devices. Two proof-of-concept designs are experimentally demonstrated in microwave region. Upon the excitation of an arbitrary linear polarization, the first device can generate distinct vortex beams with desired two linear and two circular orthogonal polarizations, whereas the second one can generate multi-foci containing components of full polarizations. This approach to generate versatile polarizations with tailored wavefront may pave a way to achieve advanced, flat and multifunctional meta-device for integrated systems.

# 1. Introduction

Recently, the emergence of metasurfaces have attracted enormous interests due to their powerful abilities to arbitrarily tailor the electromagnetic (EM) waves. By introducing subwavelength-engineered meta-atoms to produce field discontinuities across the interfaces, the intensity, polarization and phasefront of EM waves can be flexibly manipulated in desired manners [1-4]. Metasurfaces show advantages of compact size, simple fabrication, low profile and low loss compared with bulky metamaterials. Therefore, considerable amount of research works have extensively explored the metasurfaces in many aspects, hoping to fundamentally and technically fuel their development and applications as efficient and integrated devices in advanced optical or wireless communication systems [5-9]. To date, various fantastic phenomena and applications have been implemented by metasurfaces such as negative refraction/reflection [1, 2], focusing [4, 7, 10-16], super-resolution imaging [17, 18], diffusive scattering [19, 20] and special beams [21, 22], etc.

With the evolution of wireless and optical communication technologies, continuous increase of communication speed/quality, large data storage capacity, and high data throughput are urgently required. Consequently, performing several concurrent tasks or wave functionalities in a single device is especially desirable that has emerged as a promising candidate for tremendously improving the information capacity of devices [23-33]. To this goal, many efforts have been made to implement multifunctional devices through loading active components such as diodes [12, 23], graphene [24], or multiplexing techniques such as wavelength multiplexing [27, 28], polarization multiplexing [16, 29], etc. Among these methods, polarization multiplexing has its unique advantage in realizing multifunctional devices, because each polarization can be viewed as one independent information channel in the communication systems.

As one of the intrinsic properties of EM waves, polarization has been widely used in numerous applications from satellite communication and microwave detection to visible imaging and optical signal processing, all of which feature profound impacts in the modern daily life [34-39]. Ideally, a polarization-multiplexed metasurface is envisioned with the ability of being able to preserve as much as possible the incident power intensity and convert them into several independently well-defined wave functionalities in full polarization channels: orthogonal linear polarizations (LP), circular polarizations (CP). By integrating two or more meta-atoms into a unit-cell, the metasurface encoded with geometric phases can afford diverse spatial multiplexing [31] and even generate a set of full-polarized beams [32]. However, the EM wave functionalities of geometric phase metasurfaces in orthogonal circular channels are locked, due to that the phase responses for left-handed circular polarization (LCP) and right-handed circular polarization (RCP) are exactly opposite [16-18, 31-32]. This intrinsic property ultimately prevents the geometric phase from generating an assembly of versatile polarization states with independent wavefronts [32]. Needless to say, it is still rare and highly desirable to implement a metasurface capable of producing both desired wavefronts and arbitrary polarizations. This will greatly enhance the information capacity of metasurfaces and stimulate the emergence of novel compact meta-devices.

In this paper, a new approach has been proposed to overcome the aforementioned issues and demonstrated to design an integrated reflective metasurface for generating versatile wavefronts with well-defined polarization states under the excitation of an arbitrary LP. The reflected wavefront in the each polarization channel can be fully and independently controlled. We propose experimental verifications in the microwave region to validate the design principle and its feasibility. Helicity-decoupled meta-atom combining propagation phase and geometric phase is utilized to design the metasurface for generating distinct vortex beams or multi-focuses with pre-defined polarizations. The experimental results are in good agreements with the theoretical ones. More importantly, our proposed approach for realizing multifunctional metasurface is not sensitive to the direction of the incident electric-field vector, undoubtedly increasing the robustness of the polarization-versatile devices. We believe that in addition to the functions shown in the work, the salient feature of the proposed metasurface can support more fascinating multifunctional devices by employing our method.

## 2. Design method and characterization of the meta-atom

Figure 1 shows the schematic of the EM functionalities of the proposed metasurface, where multiple desired wavefronts with different polarization states can be generated from a single incident source of arbitrary LP state, and the emerging wavefront in each polarization channel can be tailored independently. To demonstrate this, we mainly rely on the meta-atoms containing propagation phase and geometric phase, and two metasurfaces with completely different functionalities are demonstrated. The first one [Figure 1a] can realize free and independent control of vortex beam generations, yielding desired topological charges and beam direction in each polarization channel. The second one can realize desired focal spots for different polarization channels with high-efficiency. Here, we define the incident polarization as $u$-LP in $uov$ coordinate, rotated by an angel of $α$ from the $xoy$ coordinate. The full polarization channels are defined as the two orthogonal LP channels of co-polarization ($u$-LP) and cross-polarization ($v$-LP), and the two orthogonal CP channels of LCP and RCP.

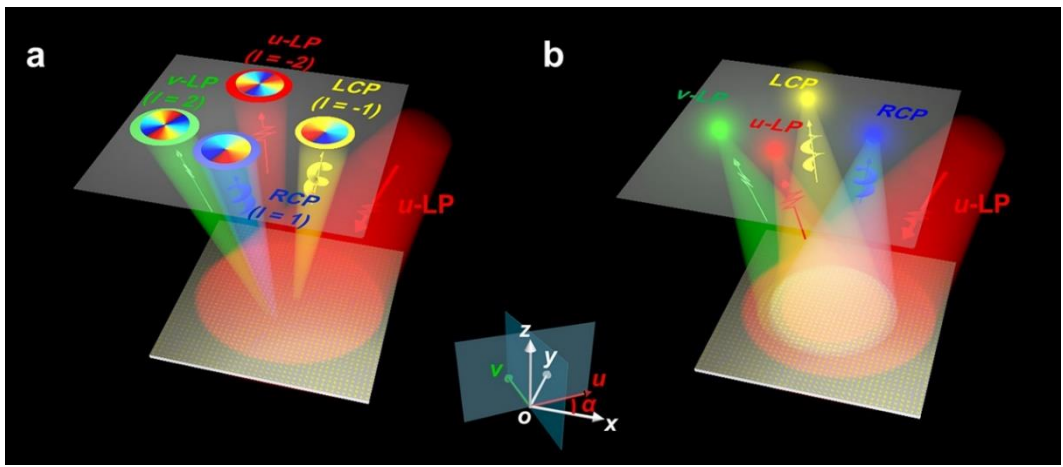

**Figure 1.** Schematic of the various functionalities of the proposed reflective metasurface with arbitrary and independent control of wavefront and polarization state under normal *u*-polarized excitation. a) The reflective beams of distinct orbital angular momentum (OAM) modes and polarization states can be controlled to arbitrary direction. b) Multiple-foci of different polarizations generated by the metasurface.

Independent control of phase profiles of the generated polarized beams is the key step for realizing complete control of the reflective wavefront and EM functionalities in each individual polarization channel. When a *u*-LP plane wave shines on a metasurface composed of finite inhomogeneous array of meta-atoms, the total reflective *E*-field from an arbitrary single meta-atom can be viewed as the collective results of the reflective e-field in each polarization channel, given as:

$$\begin{aligned}\mathbf{E}_{total}^{mn} &= \mathbf{E}_{u,out}^{mn} + \mathbf{E}_{v,out}^{mn} + \mathbf{E}_{l,out}^{mn} + \mathbf{E}_{r,out}^{mn} \\ &= \frac{1}{2}e^{-i\alpha}\left(E_u^{mn}e^{i\varphi_u^{mn}} + E_v^{mn}e^{i[\varphi_v^{mn}+3\pi/2]} + 2E_l^{mn}e^{i\varphi_l^{mn}}\right)\begin{bmatrix}1\\i\end{bmatrix} \\ &+ \frac{1}{2}e^{i\alpha}\left(E_u^{mn}e^{i\varphi_u^{mn}} + E_v^{mn}e^{i[\varphi_v^{mn}-3\pi/2]} + 2E_r^{mn}e^{i\varphi_r^{mn}}\right)\begin{bmatrix}1\\-i\end{bmatrix}\end{aligned} \quad (1)$$

where *mn* represents the position of the *mn*-th element. The parameter *α* represents the angle between the *u*-LP and x axis, which is imposed to get the desired polarization. Here, $\varphi_u$, $\varphi_v$, $\varphi_l$ and $\varphi_r$ indicate the desired phase responses in *u*-LP, *v*-LP, LCP and RCP channel, respectively. These phase profiles can be arbitrary and independent values. The $E_u$, $E_v$, $E_l$ and $E_r$ represent the amplitudes for the corresponding polarizations. Details are provided in the Supporting Materials. By independently designing the spatial phase profiles to generate desired wavefront in each channel and combining them together, we can obtain the required *E*-field reflected from the meta-atoms. Then, the problem of generating versatile polarized beams is reduced to how to realize the required meta-atoms. Actually, Equation 1 indicates that the total reflective field can be derived as a form containing two polarization components of LCP (first term) and RCP (second term), so we may conclude that if an element can realize independent phase control for LCP and RCP wave, the metasurface for generating versatile polarized beams with arbitrary wavefronts can be achieved.

For a planar reflective metasurface, the reflective characteristics of an arbitrary meta-atom placed in the *x-y* plane can be expressed by the Jones matrix as $J = \begin{bmatrix} J_{xx} & J_{xy} \\ J_{yx} & J_{yy} \end{bmatrix}$, with the first subscript indicating the reflection polarization while the second one for incident polarization. Then, the relation between incident ***E***-field ($E^{in}$) and reflective ***E***-field ($E^{out}$) of the meta-atom can be consequently determined as: $E^{out} = JE^{in}$. Here, we consider an anisotropic meta-atom exhibiting mirror symmetry along both *x*- and *y*-directions so that the off-diagonal terms in the *J*-matrix become zero. Therefore, the Jones matrix of meta-atom with merely propagation phase response is only determined by the phase responses of the two orthogonal polarizations $\varphi_x$ and $\varphi_y$. Also, we consider a geometric phase imposed to the meta-atom by assuming $\varphi_y - \varphi_x = 180°$, and then the *J*-matrix can be written as [40]:

$$J = \begin{bmatrix} \cos 2\theta e^{i\varphi_x} & -\sin 2\theta e^{i\varphi_x} \\ \sin 2\theta e^{i\varphi_x} & \cos 2\theta e^{i\varphi_x} \end{bmatrix}, \quad (2)$$

When the incident wave is *u*-LP wave ($[\cos\alpha \ \sin\alpha]^T$), the reflective ***E***-field of *mn*-th meta-atom can be determined as:

$$\begin{aligned} E_{out}^{mn} &= J^{mn} E_{u,in}^{mn} \\ &= \frac{1}{2} e^{-i\alpha} e^{i[\varphi_x^{mn} - 2\theta^{mn}]} \begin{bmatrix} 1 \\ i \end{bmatrix} + \frac{1}{2} e^{i\alpha} e^{i[\varphi_x^{mn} + 2\theta^{mn}]} \begin{bmatrix} 1 \\ -i \end{bmatrix}, \end{aligned} \quad (3)$$

which can also be decomposed into LCP and RCP components. The similarity between Equation 1 and 3 provides the possibility to realize full-channel metasurface by meta-atoms containing both propagation and geometric phase. Solving Equation 1 by addition theorem to calculate the phase responses [41] and interpreting these independent phase profiles of LCP or RCP components by the phase function shown in Equation 3, we can obtain:

$$\varphi_x^{mn} = \frac{1}{2}\arg\left\{e^{i\varphi_u^{mn}} + e^{i[\varphi_v^{mn} - 3\pi/2]} + e^{i\varphi_r^{mn}}\right\} + \frac{1}{2}\arg\left\{e^{i\varphi_u^{mn}} + e^{i[\varphi_v^{mn} + 3\pi/2]} + e^{i\varphi_l^{mn}}\right\}, \quad (4)$$

$$\theta^{mn} = \frac{1}{4}\arg\left\{e^{i\varphi_u^{mn}} + e^{i[\varphi_v^{mn} - 3\pi/2]} + e^{i\varphi_r^{mn}}\right\} - \frac{1}{4}\arg\left\{e^{i\varphi_u^{mn}} + e^{i[\varphi_v^{mn} + 3\pi/2]} + e^{i\varphi_l^{mn}}\right\}, \quad (5)$$

$$\varphi_y^{mn} = \varphi_x^{mn} + 180. \quad (6)$$

The target phase functions of each pixel of the metasurface can be acquired by the anisotropic propagation responses ($\varphi_x$ and $\varphi_y$) and the rotational angle $\theta$ (related to geometric phase) of the meta-atom. Therefore, under the illumination of an arbitrary LP incidence, the pre-designed wavefronts in all polarization channels can be generated by the metasurface if it contains both propagation phase and geometric phase. It should be noted that for each polarization we could also encode multiplexed phase profiles to generate complex beams, further enlarging the information capacity. To get the desired polarization-multiplexed metasurface, the suitable design of realistic meta-atom remains as the key point.

The meta-atom should combine the propagation phase and geometric phase to obtain independent control of the LCP and RCP wave, which requires a free control of the two phase responses for linear incidence ($\varphi_x$ and $\varphi_y$) and the orientation angle $\theta$ [38]. Such requirement can be accomplished by anisotropic meta-atoms with variable geometric parameters and orientation angle. As a design example in microwave frequency, we utilize the meta-atom as shown in Figure 2a, which has a multi-layer configuration consisting of dual-layered metallic cross-shaped resonators, a metallic ground plane, and two dielectric layers (Taconic TRF-43) with a relative permittivity of 4.3-0.0035$j$. Neighboring metallic layers are separated by dielectric spacers with thickness of $h_1 = h_2 = 1.63$ mm. The structural parameters of the meta-atom are optimized as $p = 6$ mm and $a_1 = 1$ mm and the geometries of the metallic resonator on first-layer are 0.9 times that of the one on second layer: $(a_1, l_{x1}, l_{y1}) = 0.9(a_2, l_{x2}, l_{y2})$. The parameter $\theta$ represents the spatial rotation of the meta-atom. When $\theta$ is fixed as 0° to indicate a mirror symmetry along both x- and y-direction, the cross-polarization component of the reflection wave can be well suppressed. In this case, the variations of parameters $l_{x1}$ ($l_{y1}$) can totally determine the reflective propagation phase responses $\varphi_x$ ($\varphi_y$) for LP excitation, with the reflection amplitude approaching unity at 16 GHz, as shown in Figure 2b, 2c, 2e and 2f. We can observe that for a certain polarized incidence, e.g. $x$-LP, the reflection responses are mainly determined by the parameter $l_{x1}$ while immune from the change of $l_{y1}$, which indicates a low cross-polarization talk. Figure 2d shows the reflection amplitude and phase responses when changing the length of $l_{x1}$ with fixed $l_{y1} = 2$ mm at the designed frequency of 16 GHz. Obviously, the reflection propagation phase can cover full 360° phase range while the reflection amplitude constantly retains high level (above 0.93) as the length $l_{x1}$ changed from 1 mm to 5 mm. Therefore, by adopting the proposed meta-atom and imposing a rotation operation to acquire the geometric phase, we can freely design the phase responses required for the desired functions in full polarization channels.

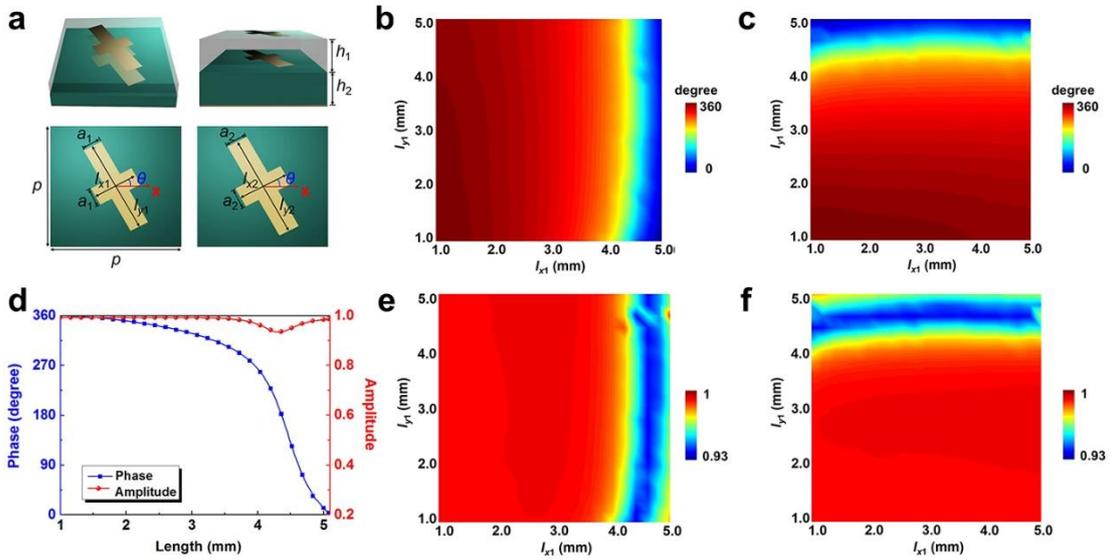

**Figure 2.** a) Schematic of the proposed meta-atom. Reflection phase responses of co-polarization for b) $x$-LP and c) $y$-LP incidence as functions of lengths $l_{x1}$ and $l_{y1}$ at 16 GHz. d) Reflection phase responses and amplitudes of the meta-atom at 16 GHz as the change of structural parameter $l_{x1}$. Other parameters are fixed. Reflection amplitudes of co-polarization in e) $x$-LP and f) $y$-LP with the change of lengths $l_{x1}$ and $l_{y1}$. The rotational angle $\theta$ is set as 0° for all cases in this figure.

## 3. Independent control of the vortex beams

We first demonstrate a practical application by utilizing the proposed meta-atom as the building blocks to independently generate desired vortex beams in full polarization channels. Vortex beams have promising prospects in future communication applications which have frequently reported by the extensive literatures. By defining the orbital angular momentum (OAM) as $l$, the azimuthal phase distribution $\exp(-jl\varphi)$ of spatial wave front is related to azimuthal angle phase $\varphi$ [42-44]. Vortex beams with orthogonal modes can simultaneously transfer at same frequency without interferences. Thus, it will greatly enhance the communication capacity without increasing the frequency bandwidth.

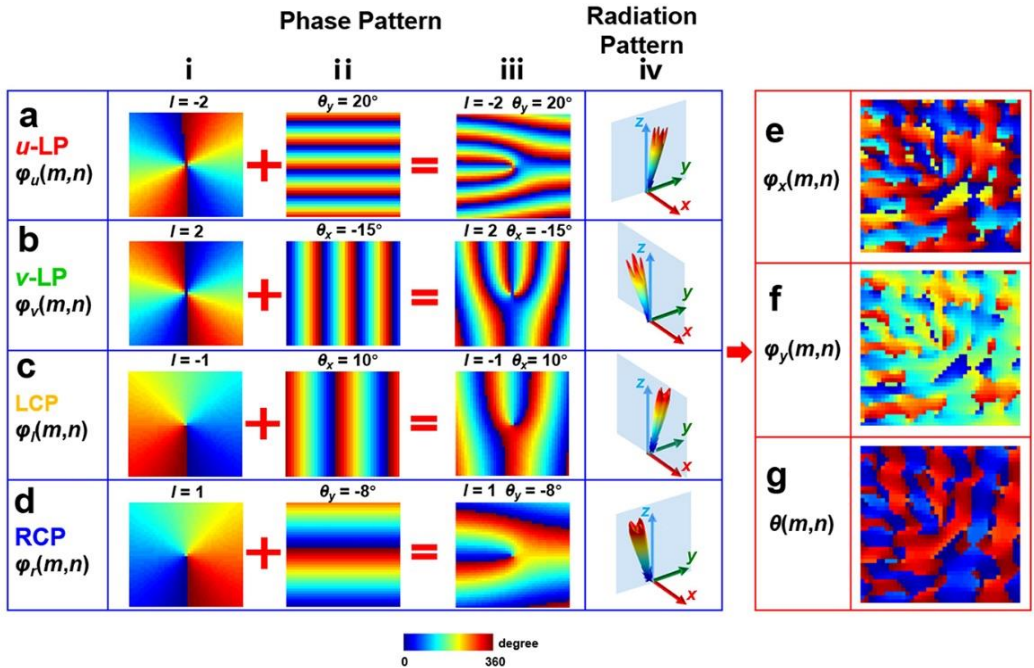

**Figure 3.** Design process to obtain the proposed reflective metasurface for independently controlling the wavefront and polarization state. The phase distributions and theoretical results of corresponding three-dimensional (3D) radiation patterns in each polarization channel: a) $u$-LP, b) $v$-LP, c) LCP, and d) RCP. Corresponding e) $\varphi_x$, f) $\varphi_y$ and g) $\theta$ calculated by the Equation 4, 5 and 6.

As schematically shown in Figure 3, we can generate vortex beams carrying different OAM modes $l = \pm 1$ and $\pm 2$ with different polarizations. The reflective beams with versatile polarizations are generated by a metasurface composed of $40 \times 40$ elements with spatially varying meta-atom structures and orientations. The designed frequency is set as 16 GHz with a normal $u$-LP incidence. First, the phase patterns with $-l\varphi$ spatial distribution to generate the vortex beams carrying different OAM modes are calculated, as shown in the first column of Figure 3. Then, to verify the design feasibility, the desired OAM beams are deflected to four different directions. In fact, the radiation direction of each designed vortex beam can be independently and arbitrarily controlled by introducing certain spatial phase gradient. By applying the constant phase gradient pattern, whose interleaved spatial phase and the lattice meets the relation of $d\Phi/dl = 2\pi\sin(\theta_r)/\lambda_0$ ($\theta_r$ is the designed deflection angle and $\lambda_0$ is the working wavelength) [1], the vortex beams can be deflected to point toward an arbitrary direction in the upper half-space. In this scenario, the applied phase gradient is shown in the second column of Figure 3, and the final phase pattern encoded in each polarization channel is shown in the third column. The fourth column of Figure 3 shows the theoretical far-filed scattering patterns generated by the encoded phase pattern in the four polarization channels. It concludes that the designed vortex beams with $l = 2$ and -1 are emitted to $\theta = -15°$ and $10°$ in the *xoz* plane, and the vortex beams with $l = -2$ and 1 are pointed to $\theta = 20°$ and $-8°$ in the *yoz* plane, respectively. Finally, to acquire the desired phase profile on the metasurface, these phase profiles in four channels are added together and transformed into $\varphi_x$, $\varphi_y$ and $\theta$ according to Eqs. (4) - (6), as shown in Figure 3e, 3f, and 3g. Therefore, we can obtain the desired $\varphi_x$, $\varphi_y$ and $\theta$ to realize vortex beams carrying distinct OAM modes with different polarizations. Interpreting these phases and orientations pixel by pixel using the designed meta-atoms, we can finally implement the metasurface for generating independent OAM beams with versatile polarizations.

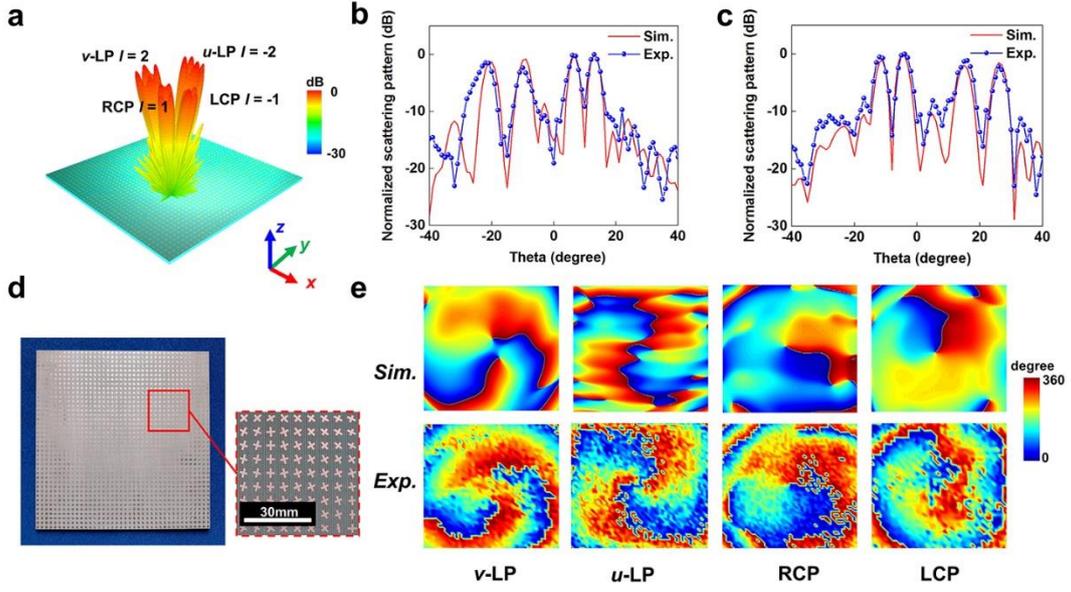

**Figure 4.** a) Simulated 3D scattering pattern of the proposed metasurface for generation differently-polarized OAM beams at 16 GHz. The simulated and measured two-dimensional (2D) scattering patterns at 16 GHz in the b) *xoz* plane, c) *yoz* plane. d) The photograph of the fabricated sample. e) The simulated and measured results of near-field ***E***-field phase distributions for the vortex beams with *v*-LP, *u*-LP, RCP and LCP states, respectively.

We perform full-wave simulations to verify the design theory using commercial software CST Microwave Studio™. The simulated three-dimensional (3D) far-field scattering pattern is shown in Figure 4a, where the vortex beams carrying the OAM mode *l* = 2 and -1 (-2 and 1) for *v*-LP and LCP (*u*-LP and RCP) with anomalous reflection angle of -15° and 10° (20° and -8°) deviated from *z*-axis in the *xoz* plane (*yoz* plane) can be observed. The simulated results are consistent with the theoretical predictions. To experimentally validate the design method and the simulations, the proposed metasurface is fabricated through standard printed circuit board (PCB) technique. The photograph of metasurface sample is shown in Figure 4d, which has a total pixels (elements) of 40 ×40 with size of 240 mm ×240 mm. Inset shows the enlarged illustration of the metasurface elements. The two-dimensional (2D) scattering patterns of the metasurface sample are measured in a standard microwave chamber and the measured results are calibrated to a same-sized metal plate. The simulated and measured normalized scattering patterns in *xoz* and *yoz* plane at designed frequency of 16 GHz are shown in Figure 4b and 4c. Because the four vortex beams have different polarization states, we measure the scattering patterns for two orthogonal LP wave, and then transform them to obtain the normalized total scattering pattern in term of power density. The vortex beams deflected by an angle of -15° and 10° deviated from z axis are measured in the *xoz* plane, and the vortex beams with deflection angle of -8° and 20° deviated from *z* axis are measured in the *yoz* plane. The measured results are well consistent with the simulated ones and theoretical predictions. To investigate the near-field behavior of the vortex beams in each polarization channel, the phase distributions of the ***E***-field of reflective wavefronts are also detected by a 3D near-field mapping system. The observation planes are set at sections perpendicular to the corresponding beam direction with an angle deviated from z-axis in the *xoz* plane (or *yoz* plane) and they are all set at a distance of $d = 800$ mm away from the center of the metasurface. The observation area is about $300 \times 300$ mm$^2$. The simulated and measured phase distributions are shown in the top and bottom panels of Figure 4e, respectively. It is recognized that the vortex beams carrying OAM with $l = \pm 1$ and $\pm 2$ can be generated in RCP, LCP, *v*-LP and *u*-LP channels. The above results validate the capability of the proposed metasurface in generating independent polarization-versatile OAM beams.

More interesting, we can observe from Equation 4, 5 and 6 that the values of parameters $\varphi_x$, $\varphi_y$ and $\theta$ to generate multiple OAM beams are immune from the change of parameter *α*, the in-plane polarization direction of the *u*-LP incidence. Therefore, we may envision that the functionalities of the proposed metasurface are not sensitive to the in-plane direction of the incident electric-field vector. To verify this, the metasurface is shined by different LP wave with parameter *α* set as 0°, 60°, 90° and 135°, as shown in Figure 5. The polarization states of the emerged wave in four channels (*v*-LP, *u*-LP, LCP, and RCP) are invariant, which clearly verify that the two orthogonal CP and two orthogonal LP waves can be well generated by the metasurface despite the change of parameter *α*.

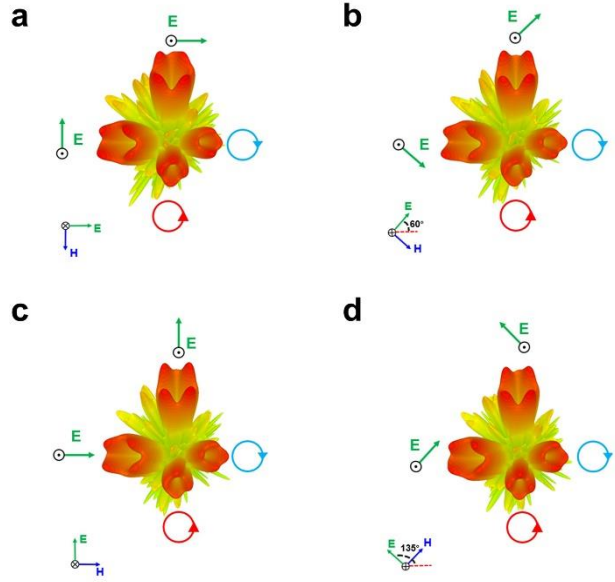

**Figure 5.** Simulated 3D scattering patterns of the proposed metasurface for different direction of incident electric-field vector (indicated at the bottom left corner) at 16 GHz: a) $\alpha = 0°$, b) $\alpha = 60°$, c) $\alpha = 90°$, and d) $\alpha = 135°$.

## 4. Generation of polarization-versatile multi-foci

Furthermore, to demonstrate the feasibility of the metasurface in controlling the EM wave built with different polarizations, in this section, we also design a metalens that can transform an arbitrary LP incident into polarization-versatile multi-foci with controllable focusing performance in individual polarization channel, as schematically shown in Figure 1b.

For a single-focus metalens, the spatial phase distribution for focal length $F$ can be calculated as [45]:

$$\Phi(x, y) = \frac{2\pi}{\lambda_0}\left(\sqrt{(x-x_0)^2 + (y-y_0)^2 + F} - F\right), \tag{7}$$

where the $x_0$ and $y_0$ represent the in-plane distance shift of the focal spot along the $x$- and $y$-axis and $\lambda_0$ is the corresponding wavelength at operational frequency. The designed metalens is composed of $40 \times 40$ elements and occupies a total area of 240 mm $\times$ 240 mm. Here, the focal length is set as $F = 8\lambda_0 = 150$ mm (or $z_0 = 150$ mm), and the positions of focal spot ($x_0$, $y_0$) in the focal plane ($z_0 = 150$ mm) are designed as (0 mm, 56 mm), (75 mm, 0 mm), (0 mm, -94 mm) and (-113 mm, 0 mm) for LCP, RCP, $u$-LP and $v$-LP focal spot, respectively. The spatial phase distributions for generating these foci are shown in Figure 6a. Then, the design process of metalens is similar to that presented above for generating OAM beams. The total phase profile for generating four-focus metalens with aforementioned focal spots in LCP, RCP, $u$-LP and $v$-Pol channels are calculated, after which this phase profile is implemented by real meta-atoms. The required profiles $\varphi_x$, $\varphi_y$ and $\theta$ to realize the proposed metalens are shown in Figure S1 in Supporting Materials. Figure 6b shows the photograph of the fabricated metalens, and the inset shows the enlarged view of the meta-atoms.

The simulated $\boldsymbol{E}$-field intensity distribution in the plane ($z = 150$ mm) at 16 GHz is shown in Figure 6c. Obviously, distinct focal spots located at (0 mm, 56 mm), (75 mm, 0 mm), (0 mm, -94 mm) and (-113 mm, 0 mm) are generated with desired polarization states. The simulated intensity distributions of $\boldsymbol{E}$-field on $xoz$ and $yoz$ plane are shown in Figure 6e and 6f. The focal spots are all located at the designed positions in the focal plane, consistent with the theoretical values. In the experiments, the $\boldsymbol{E}$-field intensity distribution on the focal $xoy$ plane are measured, as shown in Figure 6d. To give a clear comparison, we plot the simulated and measured normalized field intensities scanned along the line of $y = 0$ mm or $x = 0$ mm in the plane ($z = 150$ mm) [shown in Figure 6g and 6h]. The measured and simulated results are in good agreements and they are both consistent with the theoretical predictions, clearly verifying that focal spots with four polarization states can be generated and controlled independently.

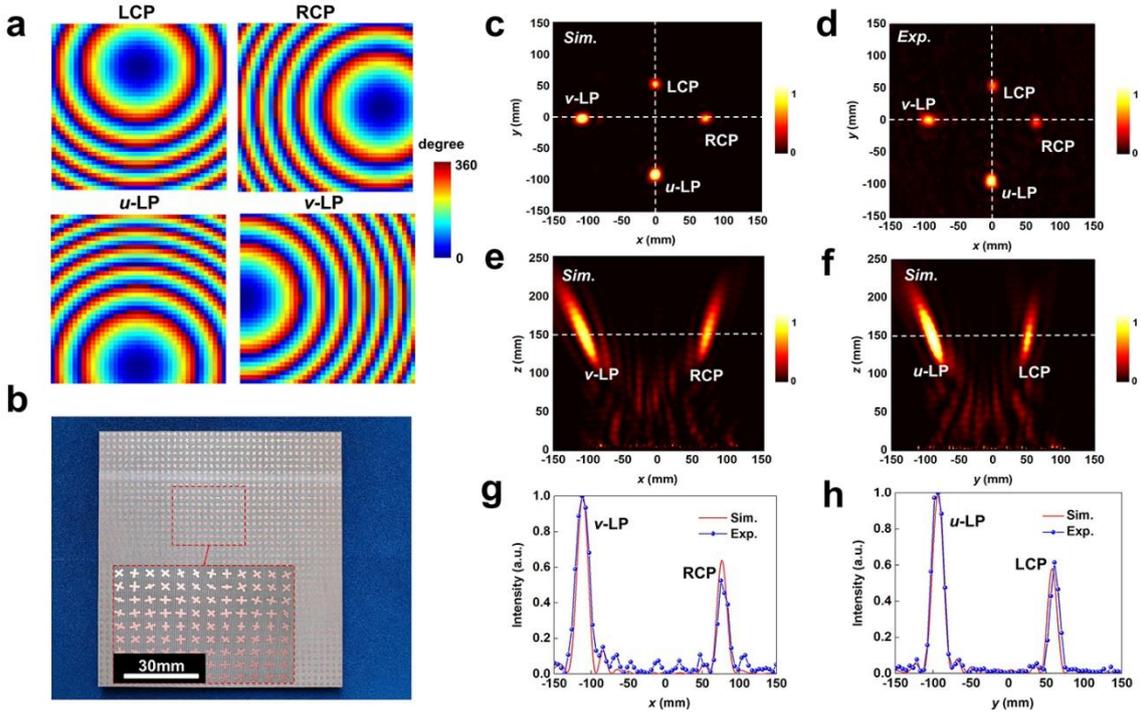

**Figure 6.** a) Phase distributions to generate focal spots with different polarizations at 16 GHz. The positions of the four foci in Cartesian coordinates are $(x, y, z)$ = (0 mm, 56 mm, 150mm), (75 mm, 0 mm, 150mm), (0 mm, -94 mm, 150mm) and (-113 mm, 0 mm, 150mm), respectively. Metasurface is located at the plane of z = 0 mm. b) The photograph of the fabricated sample. Inset shows the enlarged view. The simulated normalized intensity distributions of ***E***-field in c) *xoy* plane, e) *xoz* plane and f) *yoz* plane. d) The measured normalized intensity distributions of ***E***-field in *xoy* plane. The simulated and measured field intensities scanned along g) the *x* direction at *y* = 0 mm and h) the *y* direction at *x* = 0 mm.

## 5. Conclusion

In summary, we have proposed a method to generate full polarized beams by a single metasurface under the excitation of an arbitrary LP incidence. A general theory has been provided to guide the design of reflective metasurface to realize free and independent control of EM wavefronts with desired polarizations. In order to verify this, thin-thickness meta-atoms with low polarization cross-talk has been utilized to design two different functional metasurfaces: one simultaneously possessing distinct vortex beams with full polarizations; the other achieving polarization-versatile multi-foci. All the results are demonstrated by experiments in microwave region. More importantly, the functionalities of the metasurfaces are not sensitive to the in-plane direction of the incident electric-field vector, which will greatly enhance the robustness of these meta-devices. The design principle can be readily extended to higher frequency bands, such as terahertz and optical region. Other fascinating physical phenomena and useful functionalities can be immediately integrated as long as appropriate metasurfaces are designed/fabricated based on the proposed method. The proposed results lay a solid basis to realize polarization-versatile multi-functional metasurfaces for independently controlling the EM wavefronts in all polarization channels, facilitating the development of new meta-devices that could be applied to compact and integrated systems.

**Acknowledgements**

The authors acknowledge the funding provided by National Key Research and Development Program of China (Grant NO. 2017YFA0700201); National Natural Science Foundation of China (NSFC) (61671231, 61801207, 61731010, and 61571218). This work is partially supported by the Priority Academic Program Development of Jiangsu Higher Education Institutions (PAPD), the Fundamental Research Funds for the Central Universities and Jiangsu Provincial Key Laboratory of Advanced Manipulating Technique of Electromagnetic Wave, and the Program B for Outstanding PhD Candidate of Nanjing University.